\documentclass{ws-ijmpd}
\usepackage[super,compress]{cite}

\begin{document}

\markboth{D. Fargion, P.Oliva, P.G. De Sanctis Lucentini, M.Khlopov}
{Signals of HE atmospheric $\mu$  and traces of $\nu_{\mu}$ and $\nu_{\tau}$  in the Sun and  moon shadow.}

%
\catchline{}{}{}{}{}
%

\title{Signals of HE atmospheric \protect{$\mu$} decay in flight around the Sun's albedo versus astrophysical  \protect{$\nu_{\mu}$} and  \protect{$\nu_{\tau}$} traces  in the  Moon Shadow}

\author{DANIELE FARGION\footnote{https://orcid.org/0000-0003-3146-3932}}

\address{Physics Department \& INFN Rome1 Rome University 1, P.le A. Moro 2, 00185, Rome, Italy; Mediterranean Institute of Fundamental Physics, Via Appia Nuova 31, 00040 Marino, Italy}

\author{PIETRO OLIVA\footnote{https://orcid.org/0000-0002-3572-3255} }

\address{
Niccol\`o Cusano University, Via Don Carlo Gnocchi 3, 00166 Rome, Italy\\
Department of Sciences, University Roma Tre, Via Vasca Navale 84, 00146 Rome, Italy\\
Mediterranean Institute of Fundamental Physics, Via Appia Nuova 31, 00040 Marino, Italy
}

\author{{PIER GIORGIO} {DE SANCTIS LUCENTINI}\footnote{https://orcid.org/0000-0001-7503-2064}}

\address{Physics Department, Gubkin Russian State University (National Research
University)\\
65 Leninsky Prospekt, Moscow, 119991, Russian Federation\\
}

 \author{MAXIM YU. KHLOPOV\footnote{http://orcid.org/0000-0002-1653-6964}}  

\address{
National Research Nuclear University â€MEPHIâ€ (Moscow Engineering Physics
Institute),  Kashirskoe Sh. 31, Moscow 115409 , Russia\\
Centre for Cosmoparticle Physics â€œCosmionâ€ 115409 Moscow, Russia\\
APC laboratory 10, rue Alice Domon et Lonie Duquet 75205 Paris Cedex 13,
France}

\maketitle

\begin{history}
\received{Day Month Year}
\revised{Day Month Year}
\end{history}

\begin{abstract}
The Sun albedo of cosmic rays at GeVs energy has been discovered recently by FERMI satellite. They are traces of atmospheric Cosmic Rays (CR) hitting solar atmosphere and reflecting skimming gamma photons. Even if relevant for astrophysics, as being trace of atmospheric solar CR  noises they cannot offer any signal of neutrino astronomy. On the contrary the Moon, with no atmosphere, may become soon a novel  filtering calorimeter and an amplifier of energetic muon astronomical neutrinos (at TeV up to hundred TeV energy); these lepton tracks leave an imprint in their beta decay while in flight to Earth. Their TeV electron air-shower are among the main signals.  Also a more energetic, but more rare, PeV up to EeV tau lunar neutrino  events may be escaping as a tau lepton from the Moon: $\tau$ PeVs secondaries, then, may be shining on Earth's atmosphere in lunar shadows in a surprising way. One or a few gamma air-shower event inside the Moon shadows may occur each year in near future CTA or LHAASO TeVs gamma array detector, assuming a non negligible astrophysical TeV up to hundred TeV neutrino component (with respect to our terrestrial ruling atmospheric ones); these signals will open a new wonderful passe-partout  keyhole for neutrino, been seen along  the Moon. The lunar solid angle is small and the  muon or tau expected rate is rare, but future largest tau radio array as GRAND one might well discover such neutrino imprint.
\end{abstract}


\ccode{PACS numbers: 95.85.Ry, 98.70.Sa, 98.70.Rz}


\section{Introduction}
Neutrinos are weak interacting particles observable by their interaction with matter and by their lepton track or showers.
Because of the intrinsic interaction weaknesses a huge detector mass, such as IceCube's, is needed. The detector should screen the downward penetrating cosmic ray secondaries hiding itself in deep underground (2~km depth) Antarctic ice. Indeed,  Cosmic Rays (CR) are raining on our sky and air, creating a rich noisy signal of atmospheric pions and muons on the ground; their signals are polluting neutrino detectors. IceCube discards mostly downward penetrating muons by looking in deep underground screening ice and selecting only upcoming, detector mass contained lepton births: High Energy Starting Events (HESE) track by muons or showers by electrons, tau or Neutral Current (NC) channels. While muon tracks are well directional signal, the cascade ones are (in the ice of IceCube) smeared and almost spherical: the photons detailed shining times allowed anyway to recover a partial directionality $\pm10^\circ$. Highest energy downward muons, moreover, might be vetoed by IceTop detector array and selected only by looking at the isolated inner neutrino's births. Also through-going hundred TeV muons are used to enlarge the IceCube neutrino mass volume.

Since 2013 the discover of several highest energy ($>60$~TeV-PeVs) neutrino events (mainly shower or cascade) have suggested the injection of an astrophysical component above the atmospheric noise at TeVs energies. Indeed, the IceCube TeVs energetic muon neutrino sky showed, as expected, in the last decade to be mostly dominated by the atmospheric $\nu_\mu$, $\bar{\nu}_\mu$ noise, drowning all the eventual relevant astrophysical signals within a scrambled homogeneous insignificantly noisy sky. This map is useless like the primaries CR parents (at PeV-EeVs CR energies), which are bent by magnetic field allowing no astronomy whatsoever. TeVs atmospheric muon tracks (triggered by $\nu_\mu$) overcome by a factor $\sim20$ the electron neutrino  cascade signals; however, as it was discovered in 2013, above 60~TeV energy, the neutrino cascade (or shower) suddenly overcome by a large factor (three showers over one muon trace) the $\mu^\pm$ tracks (started by $\nu_\mu$, $\bar{\nu}_\mu$): this flavor change \cite{IceCube ICRC 2013} (or revolution \cite{2014 NIMA})  proved the rise of a new $\nu$ component very possibly of astrophysical nature.

Incidentally, it should be noticed that the  persistence  of the lack of any $\nu_\tau$, $\bar{\nu}_\tau$ signal up to now -- October 2017 -- above the detection threshold (200~TeV, nearly a dozen of events) might represent a hanging sword upon the astrophysical interpretation. Indeed, $\nu_\tau$, $\bar{\nu}_\tau$ flavors, even not produced in pion or kaon decays, if they are astrophysical, they must be present almost in equipartition with other flavors because of the expected large distances traveled and the consequent mixing during the galactic flights between $\nu_\mu$, $\bar{\nu}_\mu$ and $\nu_e$, $\bar{\nu}_e$ into $\nu_\tau$, $\bar{\nu}_\tau$. If no $\nu_\tau$, $\bar{\nu}_\tau$ will soon arise within new (foreseen twenty) or more events above 200~TeV, the most probable interpretation might (or must) be that we are thus observing mainly the injection of prompt atmospheric neutrinos of no astrophysical origin and with only comparable  $\nu_\mu$, $\bar{\nu}_\mu$ and $\nu_e$, $\bar{\nu}_e$ rate. Even in this scenario nearly a $\nu_{\tau}$ every twenty events is nevertheless expected by atmospheric charmed interactions, while nearly five in any astrophysical scenario.

As mentioned above, we have today mainly two IceCube $\nu$ skies: the downward and the upward ones with respect to the South Pole. Usually, as we mentioned, the downward traveling $\mu^\pm$ are extremely polluted by the parasite penetrating traces of hard component atmospheric air showers. The two km ice depth in IceCube do not offer enough screening to rule out those noisy $\mu^\pm$ CR tracks. These TeVs $\mu^\pm$ are nearly six order of magnitude more abundant of the $\nu_\mu$ induced muon tracks inside IceCube, or originated inside the~2 km ice depth screen. 
Nevertheless, the filtering of IceTop array  offers also a certain number of downward $\nu_\mu\to\mu$ tracks as induced neutrino events (in excess to HESE events well inside IceCube volume), being the other  signals only the up-going $\mu$ tracks whose trajectories born inside the Earth but originated outside the IceCube, anyway, they begin to suffer of the Earth neutrino opacity at energies just above the same critical $30\div60$~TeVs energy. As IceCube discovered, the flavor change occur almost exactly at those ($\sim60$~TeV) energies. Therefore the best strategy for $\nu_\mu$ astronomy is the hunt of highest energy $\mu^\pm$ tracks originated at horizon where their trajectories encounter a small cord of Earth crust and these $\nu_\mu$ might, unabsorbed, rise up to PeVs energies. These external $\mu^\pm$ induced $\nu_\mu$, $\bar{\nu}_\mu$ events enlarge the IceCube detection volume almost by a factor ten or twenty: their limited (vertical partially opaque) solid angle sky imply anyway an increase of event mass at least by a factor $\sim5$ at 200~TeV. The $\nu_\mu\to\mu$ tracks are much more directional $\pm1^\circ$ than the IceCube cascade events (whose directionality is very wide smeared being $\pm10^\circ$). Therefore, these through-going $\mu^\pm$ at highest energy has been found by us, up to now, an ideal road map to a sharp neutrino astronomy \cite{2014 Nucl.Phys.B.}, even if inconclusive results \cite{{IceCube 2015}} or rare correlation attempts with few nearby sources (within these few dozens of through going $\nu_\mu\to\mu$ tracks) were recently published \cite{Fargion Nucl.Phus.B.2017}. More attempts to find a TeV neutrino astronomy (or better anisotropy imprint)  via their cascade events map distribution has been also very recently submitted to press \cite{IceCube 2017}.

\section{The Sun's  shadow as an atmospheric muon on annular ring amplifier}

Following  earlier predictions\cite{Seckel (1991)} and later articles \cite{Ng_2016} the  bright Sun gamma albedo at GeVs up to hundred GeV has been discovered  with surprising discrepancy (underestimated) with their Fermi satellite observations and their variability anti-correlated with the sun  11 years period activity. We do not discuss here in detail the subject that has been recently searched also in IceCube and HAWC detectors \cite{2017arXiv170803732N}. We only remind that at those GeVs  up to TeVs energies the solar magnetic field plays a role in disturbing and bending charged particles; therefore we believe, the muon penetration and its thought going along the solar plasma does suffer of the solar magnetic fields bending and trapping these energy leptons. We shall address elsewhere for more details. However we expect that atmospheric solar muons skimming the Sun corona and its decay in flight  may shine at TeVs and at PeVs energy in a detectable way in future ground gamma array as HAWC and LHAASO. But these signals are mostly atmospheric cosmic ray noises as the terrestrial ones. On the contrary the Moon shadows offer a quite noise free screen to search astronomical UHE neutrino signals.

\section{The Moon Shadow as an ideal \protect{$\nu_\mu$, $\nu_\tau$} filter and screen}

The hundred MeV gamma albedo of the Moon it has been well understood \cite{Moskalenko_2007}.
The absence of the Moon atmosphere makes the higher energy gamma almost unobserved at TeV.
As one might easily note, while atmospheric  $\nu_\mu$, $\bar{\nu}_\mu$ enjoy of the high altitude (kms) Earth atmosphere to be produced by pions and muons, a much severe filter occur for CR hit on the Moon  simply because the absence of any atmosphere: a CR that hit our satellite ground, induces a ground shower where pions (or kaons for higher energies) have no much time to decay into $\nu_\mu$, $\bar{\nu}_\mu$, nor their secondaries muons dispose of many time to do so. This implies a very soft $\nu_\mu$, $\bar{\nu}_\mu$ astrophysical up-going signal from the opposite Moon face, for instance that one pointing toward our Earth, as long as regard the CR noisy secondaries. Indeed, the Moon ground might filter the downward CR noisy signals allowing the important astrophysical $\nu_\mu$, $\bar{\nu}_\mu$ to exit upward (for instance, toward the Earth) in a much softer, but clearer, spectrum band.
 Naturally a very thin annular lunar soil (ground chord of few hundreds or km size) may allow the Tevs muon secondaries to penetrate and escape toward the Earth. However this CR ground noise muons it is well suppressed because the opacity of the pions in solid lunar rocks.

\subsection{Orbital ISS arrays around the Earth or the Moon tracing upward  \protect{$\mu^\pm$}?}

For what discussed above one discover that our Moon is a much better place to hunt for astrophysical  $\nu_\mu$, $\nu_\tau$, $\nu_e$; however, it might be quite difficult to build any large twin (up-down to disentangle downward hadrons  from upward muons) area neutrino array detector inside or on the Moon ground itself. Possible large orbital stations (like ISS around  our Earth), once in orbit around the Moon might hunt much better than around the Earth for up-going astrophysical $\mu^\pm$ (although the exact energy reconstruction will be quite difficult) at $10^{10}\div10^{12}$~eV energy windows. Such an array may be dressed by scintillator plate on the top and bottom (respect the Moon side) of the hypothetical lunar ISS walls. For the present, such a detection is interesting for a ``low'' energy (several GeVs--100~Gev) $\mu^\pm$ detection associated to astrophysical $\nu_\mu$, $\bar{\nu}_\mu$. It might be possible to estimate in first approximation that such an orbital detector, assuming a detection area of 10\textsuperscript{3}~m\textsuperscript{2} along the top and bottom surface of the ideal Lunar orbital ISS, would record, assuming a 10\%  of astrophysical signals (at a tens GeV energy) respect the observed terrestrial atmospheric one, nearly an event a day or one a week. 
Unfortunately if the IceCube neutrino astrophysical spectra has a power index -2.5 the observable rate at $10^{10}$~eV will be much below of 10\%. 
For instance, assuming the observed 263 cascade events at TeV as being all of astrophysical  nature \cite{IceCube 2017}, the consequent upward astrophysical upward $\mu^{\mp}$ by $\nu_\mu$, $\bar{\nu}_\mu$ at TeV, will be increase (for a 10\textsuperscript{3}~m\textsuperscript{2} ISS detector) by a factor (due to muon penetrability at TeV and a rock lunar density)  $\simeq10\cdot10^{-3}$ leading to  $\simeq2.63$ event a year. Eventual large size underground array should be built (as large as IceCube) if the Moon had ice, as discussed below for other Icy Moons. But our Moon has not icy soil.
In conclusion as we already know by the IceCube DeepCore (and by Super Kamiokande early data) $\nu_\mu$, $\bar{\nu}_\mu$ at these GeV$\div$100~GeV energies are ruled by atmospheric noisy signals. Also Lunar signals ones at GeVs energy might be partially polluted by cosmic traces. Therefore such a spectacular large array ISS in orbit around our Moon, while being important for general future science test, it seem an extreme costly neutrino experiment by a dubious and a poor astrophysical $\nu$  benefit.

\section{Orbital Cherenkov telescope observing upward cascade flash in orbit over Icy Moon}

There is a very  easy way , in far future, to track up-going neutrino signal rate: the cascade on Icy Moon, made by charged current $\nu_{\tau}$, $\nu_{e}$  cascades in exit from the Moon, as those PeVs showers observed in IceCube. They should also shine, while showering on surface, lights outward  the Icy Moons external surface toward orbital satellites. Because the absence of the atmosphere (excluding Triton moon) there are not disturbing downward air-shower CR on Icy Moon, downward air-showers whose shining (observed from high altitudes) might be confused as an upward neutrino cascade. There should occur that vertical CR or UHECR may produce in ice cascades that anyway will increase and shine down deeper and deeper in Icy Moon ice: their random walk photons may be resurgent as up-going cascades; however their timing spread and structure should be well disentangled from neutrino signals.
Smallest Icy Moons (few hundred km size) may be totally transparent to upward PeV or even EeV neutrino energy.
Of course this possibility should be better  estimated in detail once each Icy Moon in the solar system will be probed
either for its ice  density, albedo and especially for its internal ice transparency. For a first approximation any (Jem-EUSO like) telescope in orbit at 40~km. altitude from the Icy Moon might be able to reveal any surface  up-going ice-shower or cascade within its angle view of $45^{\circ}$ toward a down icy moon, neutrinos at energies above or comparable of PeVs or tens PeVs. These skin outward cascades will simply lit the surface and  eject  their thousand billion  Cherenkov photons: they may reach (at least a several dozen or even fifty ones in few tens nanosecond) at orbit detector, being recorded by a meter size Cherenkov telescope.
The mass equivalent observable by such PeVs lightening on the Icy Moon is just, let say, associated with a ten meter depth of the ice (its assumed transparency) and with a disk area (from 40 km altitude) of nearly 5000~km\textsuperscript{2}. Therefore the target mass equivalent for such up cascade moons is comparable at least to 50~km\textsuperscript{3} w.e. (water equivalent), just fifty times IceCube mass leading to a rate of expected one event observable every  week.
One cannot exploit such a mining photon system easily while orbiting around our Earth because of the Earth vertical opacity to neutrinos at PeVs  (already quite severe), and because the needed satellite distance from the Earth  surface (400~km or more distance), it makes difficult such a detection. Moreover, the presence of a huge rate of downward PeVs-EeVs air-showers makes very noisy and difficult for the orbital detection of such up-going flashes by ice-shower on Earth (ice or even, just in case, quite lake sea surface). Anyway the rate of upward tau air-shower, somehow comparable at PeV to these skin cascade has been considered long time ago \cite{Fargion_2004}.
In conclusion Jew-EUSO like detector in near Icy Moon orbit maybe an ideal count rating of upward PeVs escaping neutrino shower. Unfortunately the directionality of such events are difficult as in IceCube or even worst: therefore the Icy Moon calorimeter cascade role is not optimal for any neutrino astronomy but for a larger spread anisotropy study  and for a better neutrino statistical counting. Finally we wish to mention that the most far (from the Sun) Icy Moon  will suffer of the less solar day-light planetary reflected light noise. Therefore mini Icy Moon as the Pluto ones, for instance mini Nix, whose albedo is as bright as a perfect snow (Icy~II) it might be one of the best candidate target for its Moon shadows in high energy neutrino astronomy.

\subsection{Glashow resonance \protect{$\bar{\nu}_e + e$} upward cascade flash outside the Icy Moons}

One of the most remarkable signal to be detected by upward neutrino skin cascade is the enhanced Glashow resonant interaction: $\bar{\nu}_e + e\to W^{-}$; indeed it may outcome and interact on icy surface (a few tens meter soil) while being crossing along few hundreds (or even one thousand) km diameter Icy Moon size. Indeed such a small size Moon is still transparent to the such very interacting  resonant reaction, whose interaction length is of few hundreds km length; the resonant energy $E_{\bar{\nu}_{e}} = 6.3$~PeV is even able, some times (17\%), to trace a tau that it may shine, while in later decay, once again by a flash: first by a direct boson $W^{-}$ birth, than by its tau lepton decay few, $\simeq300$ meter $\tau$ decay distance increasing the probability to be observed at the surface.
This peculiar double bang channel may be as a key astrophysical neutrino trigger \cite{Learned_1995}, up-today, unfortunately, unobserved \cite{Fargion_2016b}. The first bang inside the matter and the second one, outside in air, is the base of the well known tau air-shower (or Earth skimming) way to tau neutrino astronomy \cite{Fargion(2002)}.   The cascade lights penetrating size (a few tens meters) versus the resonant neutrino distance is nearly $\leq10^{-4}$, making such a signal quite probable.
At the present IceCube didn't yet reveal this important event. As we mentioned above such soil cascade lightening on Icy Moons are almost spherically symmetric leading to a poor directionality message, but a high detection rate: a very promising neutrino astrophysics test that might be located  in most wide mission (as the biological life search toward Enceladus) in any future mission along any Icy Moons.

\section{Our Moons shadows on Earth as filter for several tens TeVs \protect{$\nu_\mu$}}

Let us briefly mention that the idea of using the Moon for the UHE neutrino detection is not new at all.
Since the brightest idea of Gurgen Askaryan  to look for coherent charge separation inside the matter
while internal shower occurs and to search their radio emission \cite{Askariyan (1962), Askariyan (1965), Askariyan (1979)}  there have been a large chain of proposal confirmations and also very exciting on going experiments
\cite{Gorham 2016}: most of them are looking to the Moon for the extreme UHE neutrino at ZeV $10^{21}$~eV energies able to shine Moon while skimming and  showering into radio flashes along the Moon corona edges.
The Askaryan effect play a role mainly at those highest energy, but it is not of our present interest: indeed
our proposal is more focused below and above PeVs  $10^{15}$~eV  energies already observed in IceCube; it may also extend up to $10^{18}$~eV energy of GZK interest as shown later.

Let us now consider the very realistic case of an ultra-relativistic  $\nu_\mu$, $\bar{\nu}_\mu$ crossing the Moon at TeV energies pointing the Earth: its possible interaction within the Moon surface skin may lead to an escaping $\sim$ TeV muon whose flight is still toward our planet Earth. Now a unique muon track  (within the Moon shadow) is mostly useless, even it maybe in principle detected in IceCube, but its amplified secondaries are not. These rare gamma like air-shower are of lunar origination and they may spread in wide areas (Moli\`ere radius of hundred meter). Unfortunately any CR bending angle due to the solar magnetic field along the trip Moon-Earth might be able to deflect somehow all charged particle (of charge $Z$ and energy $E$)  by a quantity
\begin{equation}\label{eq:1}
\Delta\theta\simeq1.6^\circ\left(\frac{E}{1\,\mathrm{TeV}}\right)\cdot Z
\end{equation}
This CR nuclei as well any proton or electron signal may pollute at this energy the dark Moon shadows.
Incidentally this pollution of positive and negative protons allowed first bound on anti proton spectra since long time ago.
Indeed the Moon radius angle size it is much smaller than the bending one above,  as observed from Earth:
\begin{equation}
\Delta\theta_{\mathrm{Moon}}=\arctan{\left(\frac{R_{\mathrm{Moon}}}{D_{\oplus-\mathrm{Moon}}}\right)}\simeq0.26^\circ
\end{equation}

Consequently, at energies above $E_{\mu^\pm}\gtrsim 6.4$~TeV, the bending due to magnetic Lorentz force is contained inside the Moon shadow and thus as a first approximation it may be neglected any (Z=1) CR pollution. There are anyway heavier  and larger charge nuclei whose bending may bend in lunar disk, making an hadronic air shower noise, but they are not electro magnetic noise at all. Therefore such a imprint maybe revealed.
 Let us notice that the Moon solid angle with respect to the whole sky is just a tiny fraction:
\begin{equation}
\frac{\Delta\Omega_{\mathrm{Moon}}}{\Delta\Omega_{4\pi}}\simeq5,3\cdot10^{-6}
\end{equation}
therefore if we would estimate, as a very first approximation, how  would be in IceCube, the energetic (TeV) muon sky, $\nu_\mu$-induced, or HESE within this thin Moon Shadow, we would find a quite rare event rate. Indeed, at the present IceCube observes nearly $10^5$ event per year of up-going muons from $2\pi$ section of the sky. Therefore, the consequent event rate would be just $\sim2\cdot10^5\cdot5\cdot10^{-6}\sim 1$ event per year, coming from the Moon shadows (neglecting the bending discussed above as well as the Moon's opacity which results negligible at TeVs or at hundred of TeVs energy band).
 Naturally we deal with a Moon shadows while the Moon is on the same side of the detector sky; the case of the Moon on the opposite Earth side is now just of academic interest and it is neglect.

\subsection{Muons decay in flight feeding the unique neutrino induced TeVs electromagnetic air-showers}

Now we consider the consequent imprint due to such a TeVs (better to say, for unbent signal, $E_{\nu_\mu}\gtrsim 6$~Tev) $\mu^\pm$ born at Moon and on their arrival way to Earth. Let's remind that the decay time of a muon is $\tau_\mu=2.197\cdot10^{-6}$~s and that the muon mass is $m_\mu=105,65$~MeV, in order to estimate the distance life in flight of an ultra-relativistic $\mu^\pm$:
\begin{equation}
L_{\mu}\simeq6234.26\,\mathrm{km}\;\left(\frac{E_\mu}{1\,\mathrm{TeV}}\right)
\end{equation}

As we know the average distance Moon to Earth is about $D\sim384400$ km. Thus at energies between 6.4~TeV and $\frac{D}{L_\mu}\simeq 61.66$~TeV the muon will linearly decay as an electron in its flight to us, shining inside the dark Moon Shadow. These produced $e^\pm$  will then hit the Earth's atmosphere causing a very rare and amazing electromagnetic airshower around or from inside the Moon's disk, TeVs photons whose presence is already extremely rare in all the gamma TeV skies. How many of such $e^\pm$ signals originated by a grandparent TeVs $\nu_\mu$, $\bar{\nu}_\mu$ in the Moon external ground disk would reach our planet? To make a first estimation of the probability $P_{\nu_\mu\to\mu}$ to be originated on the Moon we need to know the ratio of the muon track length respect to the UHE $\nu_\mu$, $\bar{\nu}_\mu$ interaction lengths.

Analogous arguments were extensively discussed in details for $\nu_\tau$, $\bar{\nu}_\tau$ and $\bar{\nu}_e$ produced inside the Earth crust and escaping as $\tau$-airshowers, \cite{Fargion(2002), Fargion_2004, 2007NuPhS.165..207F, 2007NuPhS.165..116F, 2007NuPhS.168..292F}. Here we briefly remind that at energies above $30\div60~$ TeV the Earth diameter, comparable to $\sim10^{10}$~cm water equivalent, starts to become  opaque to UHE neutrinos. The smaller Moon diameter slant depth (being on average less dense of Earth) is $\sim8\cdot10^9$~cm w.e., therefore the Moon begin to represent an opaque volume to UHE$\nu$ at $E_\nu>2$~PeV. Thus, up to energies
\begin{equation}
6.4\;\mathrm{TeV}\lesssim E_\mu\lesssim 61.66\;\mathrm{TeV}
\end{equation}
there is not much direct $\nu$ opacity throughout the Moon's volume: this implies no $\nu$ Moon absorption  up at least to a few PeVs. Moreover, the Moon crust is made by rock whose density is twice and half more dense than water (or ice); as a third remark, the external crust where UHE $\nu_\mu$ (at least by several TeVs energies) may interact and its $\mu$ track may cross along, may exceed  $10\div30$ km distances. Therefore, as a first approximation the $\nu_\mu\to\mu$ event track length penetrate  nearly (or more) than 25 km w.e. with respect to the same one event inside the 2 km w.e. above IceCube or just the one km. for HESE events. As a simple consequence we may estimate, as done above, just one event per year of this kind $\nu_\mu\to\mu$ in HESE inside IceCube toward the Moon, as well as two downward muon event per year made by the 2 km of ice height (that is born above IceCube), as well as we estimate nearly 25 events downward per year of TeV $\nu_\mu\to\mu$ for square km. escaping from the Moon (rarely also toward Earth).
These TeVs $\mu^\pm$ flight do not reach us because their decay in flight, but their decay products, the $e^\pm$, will survive and eventually hit the atmosphere, twice a week for square km, shining as $\gamma$ air-showers. Future CTA array gamma detectors (whose energy threshold best sensitivity is at best at TeV energy), might reveal such signals around (because of partial bending) or inside the Moon Shadows. The advantage of such a signal is the amplified (factor $\simeq 10-30$) ability of the Moon to make muons and the absence of atmospheric noises. Naturally several TeVs CR, light or heavy nuclei, bent inside the Moon shadows may make a huge noises, but they are different  from any electromagnetic air-showers; therefore they may be disentangled by their shower morphology in CTA or other array detectors.

As we underlined in \eqref{eq:1}, these TeVs Moon muons bending make the eventually reconstructed arriving angle trajectory pointing outside the Moon disk at TeV but at energies above $E_{\nu_\mu}>6,4$~TeV these neutrino originated $\mu^\pm$ will probably hit from a direction pointing inside the Moon disk. Of course we assume here a comparable $\nu_\mu$ flux on the Moon as the Earth one. A more accurate estimate should consider not the atmospheric but just only the astrophysical fluxes at TeVs.
 We could ask ourselves if we may observe these several events each year of $e^\pm\to\gamma$ airshower signals coming from the disk of the Moon and around it. We remind  the reduced TeV $\gamma$ background noise in the sky and we remind the favorable  present detectors (and the near future ones)  gamma threshold. For instance, HESS Cherenkov minimal fluency detection at $\sim$TeV energy is $\sim0,1$~eV~cm\textsuperscript{-2}~s\textsuperscript{-1} \cite{2014arXiv1403.4550C} while the future Cherenkov Telescope Array   (CTA, a km\textsuperscript{2} array) would reach 10\textsuperscript{-2}~eV~cm\textsuperscript{-2}~s\textsuperscript{-1} \cite{2016arXiv161005151C}. For the
The  Large  High  Altitude  Air  Shower  Observatory (LHAASO)  project  the array will encounter the best threshold at 100 TeV reaching $6,2\cdot10^{-3}$~eV~cm\textsuperscript{-2}~s\textsuperscript{-1} sensitivity \cite{2016arXiv160207600D}.
This implies that if the conversion $\nu_\mu\to\mu$ has a probability of 10\textsuperscript{-4}, as much as it results at $\simeq10^{13}$ eV, than the signal ($\sim30$~eV~cm\textsuperscript{-2}~s\textsuperscript{-1}~sr\textsuperscript{-1}) would be at detection threshold $3\cdot10^{-3}$~eV~cm\textsuperscript{-2}~s\textsuperscript{-1}~sr\textsuperscript{-1}. However, the Moon solid angle is very small indeed ($\Delta\Omega_{\mathrm{Moon}}\sim5\cdot10^{-5}$~sr) making the rate any way quite rare.

Once again we underline that the rate of the observed HESE events in IceCube is related to the interacting km\textsuperscript{3} ice mass, while for the same fluency ($\sim30\div60$~eV~cm\textsuperscript{-2}~s\textsuperscript{-1}~sr\textsuperscript{-1}) of these $\nu_\mu$ originated on the Moon, they will
escape from the solid lunar skin crust in an increased muon track mass in its way out. Thus, the energy fluency of such TeVs $\mu^{\pm}$ (as well as their number flux) is increased by a factor related to the ratio between the Moon's rock density and the water one $\rho_M/\rho_w\simeq2,5$, and of course related also to the additional distance depth ratio traveled by TeVs muons $L_\mu/L_{\mathrm{IceCube}}\sim20$, in analogy to trough going muons, induced by UHE neutrinos in IceCube.

The over all consequence is that the effective fluency of $\mu^{\pm}$ due to 10\textsuperscript{13}~eV $\nu_\mu$ crossing the Moon and possibly shining (as secondaries $e^\pm$) inside the Moon Shadow  (shining as a gamma airshower) is
\begin{equation}
\phi_{e^\pm}\approx3\cdot10^{-3}\cdot50\simeq0,15\,\frac{\mathrm{eV}}{\mathrm{cm}^2\,\mathrm{s}\,\mathrm{sr}}.
\end{equation}
The final expected flux in energy from the Moon solid angle becomes
\begin{equation}
\phi_{e^\pm}\approx1,5\cdot10^{-1}\cdot6\cdot10^{-5}\simeq10^{-5}\,\frac{\mathrm{eV}}{\mathrm{cm}^2\,\mathrm{s}}.
\end{equation}
In comparison with  LHAASO array detector threshold this number is, at its pessimistic approximation, nearly 600 time smaller, therefore one may suggest an even a wider (but diluted every 20-30~m) array in a much wider area, as the Telescope Array one ($\sim$600 km\textsuperscript{2}) which will be able to observe $\nu_\mu\to\mu^\pm$ of astrophysical origin. Such an array area may well be comparable  to Telescope Array (TA) and even below Auger huge areas ($\sim3000$ km\textsuperscript{2}). However our suggestion to open a window toward the Moon for muons decay signal has a key feature and advantage: those UHE $\nu_\mu\to\mu^\pm$ at TeVs-hundred TeVs are totally free of any atmospheric noise, simply because the Moon has no atmosphere. Only prompt (charmed) lunar atmospheric neutrino may also inject rare $\nu_{\mu}$  noisy signals. But very rarely at tens TeV energy and not (almost at all) for $\tau$ charmed flavor, that moreover cannot oscillate from $\nu_{\mu}$ at those distances and energy even from the Moon. Therefore  these electron TeVs airshowers inside the Moon Shadows are a very new exciting road to neutrino astronomy, even now quite rare. One or a few gamma air-shower event in Moon shadows each year in CTA or LHAASO array detector may be a new wonderful passepartout via keyhole of the Moon.

\section{Hundred TeV Muons tracks from the Moon into the IceCube}

The extended IceCube at tens square km. area might be able to reveal also rare deep penetrating muons originated on the Moon and so energetic to not yet being decayed. The eventuality that such a hundred TeV neutrino was on the contrary born on Earth ice (above or inside IceCube) is negligible: few $\simeq 10^{-2}$. However such an event rate is anyway poor. In an ideal ten-hundred IceCube detector where several million of event a year maybe collected there will be in a few years somehow a visible  over crowding of (downward) hundred TeV muons (made by  lunar neutrinos)  within the Moon shadow disk, making its hundred TeV sky brighter  than else where. Finally
let us remind that among the eventual  future radio array for tau neutrino air-shower, the recent  proposed GRAND (The Giant Radio Array for Neutrino Detection) \cite{2016_GRAND} is extremely vast, as wide as $2 \cdot 10^{5}$ $km^{2}$ tuned to tens PeV $\tau$ signals from Earth or Mountain chains, the same experiment might be well finalized also to such a $10^{14}$ eV radio signals from the Moon. In that case the rate of observable muons from the Moon maybe a really huge number, a few a day.
In the  PeV energy in what it follows, we want to mention also a higher energy $\nu_\tau$, $\bar{\nu}_\tau$  surprising imprint.

\section{UHE PeVs and EeVs $\tau^{\pm}$ escaping from the Moon surface skin and their pion relics decaying (or not) to Earth}

The $\nu_\tau$, $\bar{\nu}_\tau$ neutrinos do produce escaping associated lepton $\tau$ with a much shorter (at PeVs energies) length $L_{\tau}\simeq 49 m. \frac{E_{\tau}}{PeV}$ respect to $\nu_\mu$, $\bar{\nu}_\mu$ (as long as tens km length, growing in logarithmic law with energy). PeVs $\nu_\tau$, $\bar{\nu}_\tau$ energetic $\tau$ air-shower  signals (very useful when one wants to look for them in terrestrial deep valleys) are not crossing deep distance nor they are leaving much of a trace from the Moon. Anyways, as their energy reach 10\textsuperscript{17}-10\textsuperscript{18}~eV, the $\tau$, $\bar{\tau}$ lepton tracks become comparable and then even longer of the ones produced by the more conventional lepton $\mu^\pm$. This implies an important competitive mechanism in these energies  toward Earth from Moon Shadow.

\subsection{The \protect{$\pi^{\pm}$} and  the \protect{$\pi^{0}\mapsto 2 \cdot\gamma$} tracing \protect{$\tau^{\pm}$} , decay in flight toward Earth}

At EeV energy the lepton $\tau$ is able, by Lorentz boost, to fly in vacuum at least 49 km (at 1~EeV) while inside the solid rock it may reach several tens of km, even further than a $\mu^\pm$. The $\tau$ decay in flight might offer very interesting signature, in connection with its main different secondary imprint (hadronic, such as $\pi^\pm$ and electromagnetic via $\pi^0\to2\gamma$). Mostly ($67\%$), $\tau$ decay happens in hadronic channel. Let us remind that the pion lifetime is  $t_\pi=2,6\cdot10^{-8}$\,s, then for a pion with a rest mass of $m_{\pi^\pm}=139,57$~MeV the travel  distance is as long as
\begin{equation}
L_{\pi^\pm}=55918\cdot\left(\frac{E_{\pi^\pm}}{\mathrm{PeV}}\right)\,\mathrm{km}
\end{equation}
so that, being the Moon distance nearly 7 times longer, there is a bound on the $\pi^\pm$ energy giving a range of
\begin{equation}
6,87\,\mathrm{PeV}\gtrsim E_{\pi^\pm}\gtrsim 61\,\mathrm{TeV}
\end{equation}
Below this energy pion is not allowed  to reach us form the Moon (while of course their $\mu^\pm$ will reach our planet).
Then, above this upper limit energy ($\sim7$~PeV), $\pi^\pm$ will travel and hit themselves the Earth producing a very `` surprising and otherwise unexplainable''  shower with an hadronic signature inside the normally ``silent'' Moon Shadow. No heavy nuclei at 7 PeV may be systematically bent in such a way to mimic and to appear as they are coming from the Moon disk.

\begin{figure}[pb]
\centerline{\psfig{file=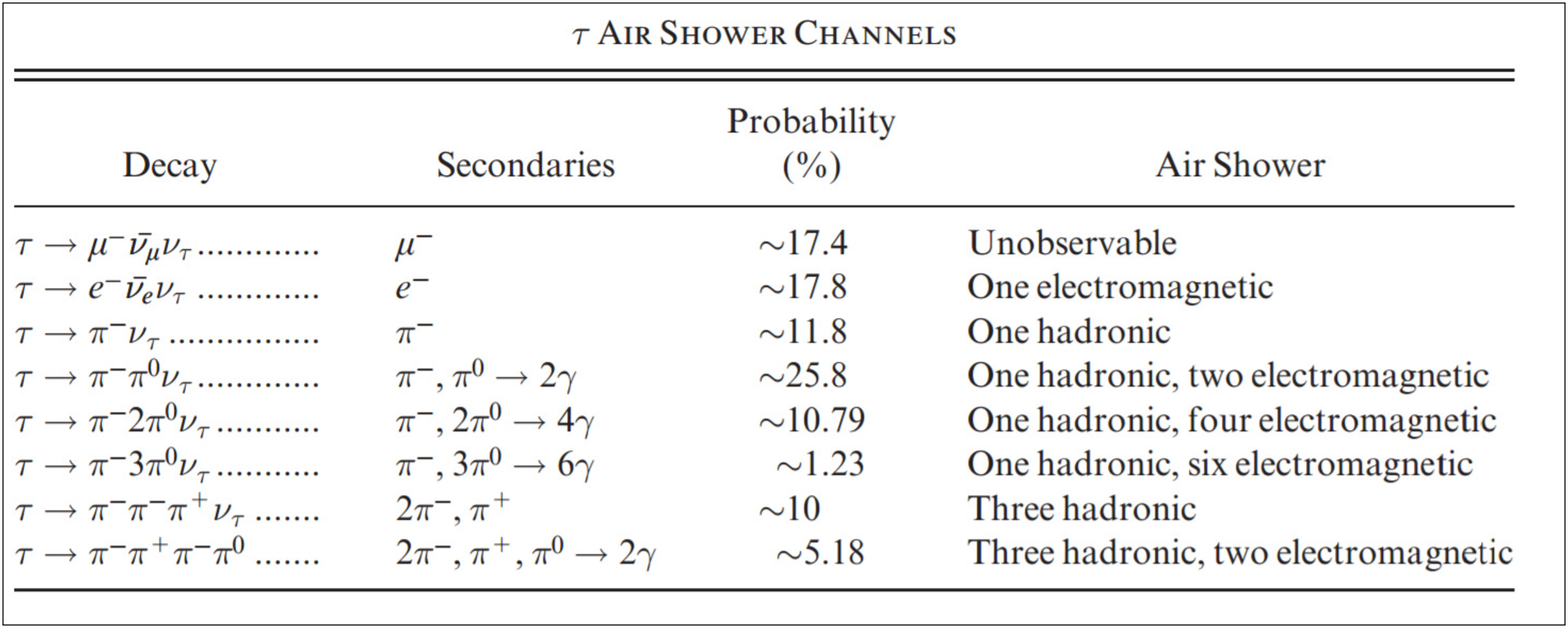,width=11.0cm}}
\vspace*{8pt}
\caption{The ultra-relativistic PeVs-EeVs  tau decay channels may lead to a ramification of decays whose secondaries showering at Earth on terrestrial atmosphere lead to different surprising electromagnetic and hadronic shining inside the Moon Shadow \cite{Fargion(2002)}.}\label{fig:2}
\end{figure}

An additional ``otherwise unexplainable'' key signal will be a formidable a PeVs twin photons, due to the prompt neutral pion decay from tau secondary tail, reaching us from the Moon Shadow at once. Indeed, the $\pi^0\to2\gamma$ at 1,4~PeV (possibly out of a few PeV $\tau$) energy will produce  a boosted system where their two  photons will appear to us almost in parallel trajectories ($\Delta\theta\sim m_\pi/E_\pi\sim10^{-7}$~rad).
Therefore, there may be a very unique imprint of a twin $\sim$ 700 TeV gamma air-shower at a distance of nearly 40 meters or more from each other due to such a rare UHE $\nu_\tau\rightarrow...+\pi^0$ (see Fig.\ref{fig:2}) neutral pion decay channel, both coming inside the dark Moon shadows. What makes exciting these Moon rare (neutrino induced)  signals are the possibility to be revealed also in foreseen GRAND experiment where not only the Earth crust (observed by radio antenna array on top mountain location) will be the tens PeV $\tau$ neutrino calorimeter, but also the same Moon may tracks its EeV or tens PeVs $\tau$ decay in their wonderful  and rich nuclear physics  traces.
\subsection{The multi tree pions and photons decay in flight toward Earth}
The following table it does show the complex channel tree of the tau decay in laboratory and in flight. It must be important above PeVs energies where the $\tau$ track inside the Moon made the tau escape more and more probable. As we already mentioned and as it is shown in the table there are 17.4 \% probability to have an unique observable secondary muon that will not decay but it may penetrate (in extended) IceCube detector. There are also $17.8 \%$ probability to have an unique UHE electron making  also an unique electro-magnetic (El.Ma.) air-shower. There  is the $11.8 \%$ probability to observe an unique (pion survival) and hadron shower, inside the Moon Shadow.
There are more probable and surprising 25.8 \% probability to observe one hadron and two photons leading to two aligned El.Ma. airshower, almost all (at PeVs) in a few hundred meter size area. In analog, more surprising a cluster of one hadron and four El.Ma. airshower at 10.79 \% probability; additional tree hadron cluster 5.18\% probability may also occur with two El.Ma. airshower  or without El.Ma. airshower at 10 \% probability rate. In conclusion the Moon shadows might hide the inner nuclear $\tau$ physics in a very spectacular way. The eventual GRAND \cite{2016_GRAND} array radio detector might be able in near future also  to discover (within a much rich upward terrestrial tau signature) also these rare but spectacular events, each once or  several times a year.

\begin{figure}[pb]
\centerline{\psfig{file=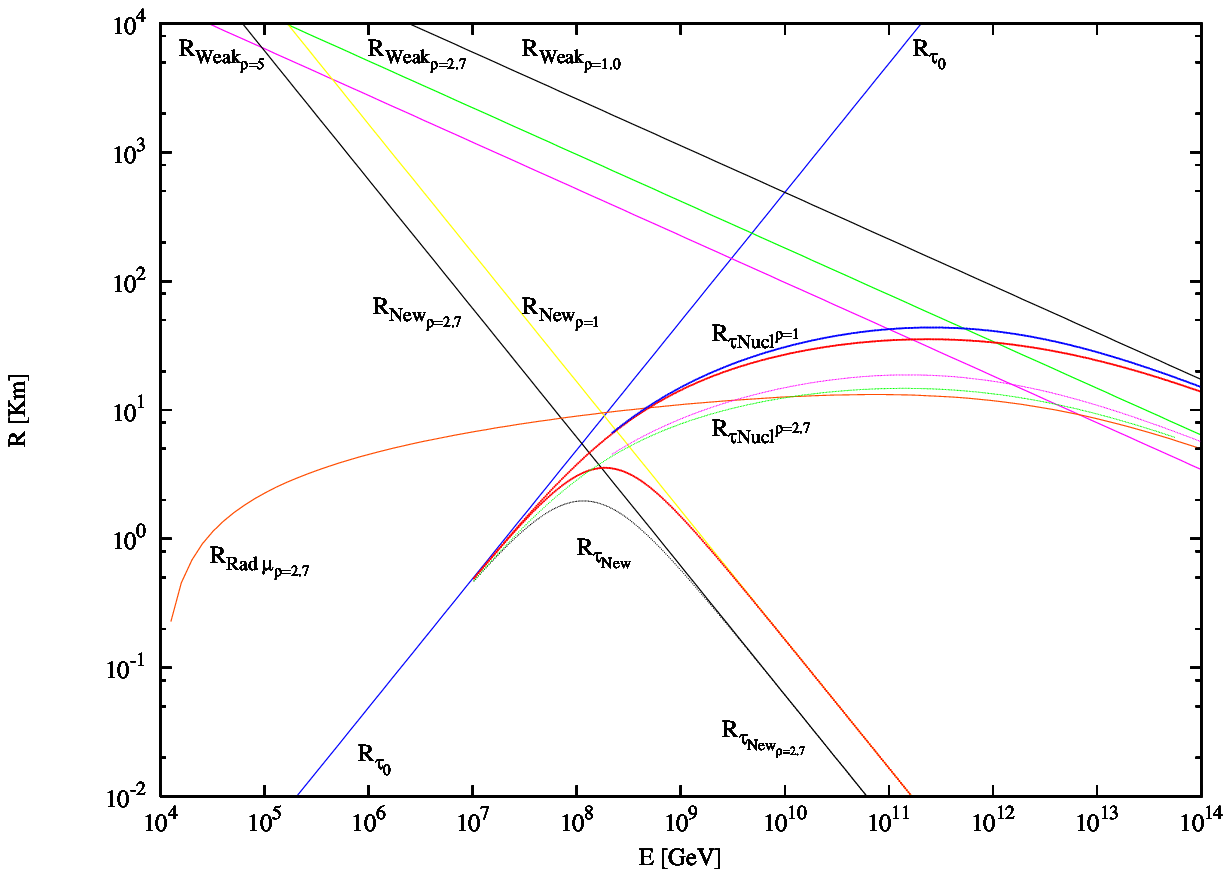,width=11.0cm}}
\vspace*{8pt}
\caption{At ultra-relativistic regime $\tau$  reach and overcome muon tracks though in less extreme conditions muon lepton is the more penetrating one. The tau indeed offers an interaction length which is derived by an hybrid transcendent equation linking energy losses and life-time length  \cite{Fargion(2002), Fargion_2004}.}
\end{figure}\label{fig:1}


\section{Conclusions: Muon and Tau neutrino traces in Moon disk}
We are aware that the solar shadows and its albedo are an important and brightest test for the gamma sky astronomy and for the cosmic ray traces in solar plasma. However there are several new and surprising (less bright) signals arriving from the Moon as a huge calorimeter for UHE neutrino to the Earth.
The analogous Askariyan \cite{{Askariyan (1979)}} radio signal  is  a competitive trace  at ZeV energies, but the neutrino astronomy we considered here it is at PeVs energies, a more probable and frequent astronomy at IceCube edges,
both for muons and  (at a much higher energy) for tau  (GZK) neutrinos. The complexity of the signal is rich and cannot be  neglected. Particle physics and air-shower physics are building this new windows of $\nu$ astronomy via Moon Shadows. The GRAND hundred thousand sq. km detector array in project (or even just the sq.km detector as LHAASO) maybe already at the detection threshold (by several event a year for an astrophysical neutrino flux comparable ($20\%- 5\%)$) with the terrestrial atmospheric ones). Better rates might reach from very extended CR arrays (as AUGER or TA extended at tens PeV energy range).
 There are several marks, as the TeVs and the hundred TeV muon neutrinos, the Glashow resonance fine tuned band, the tau decay channels and their pion decay in flight: all of them make very bright this novel independent road to the high energy neutrino discover. The solar secondaries albedo in gamma sky are overcrowded by CR solar atmospheric noises even larger than the terrestrial atmospheric ones and much more than lunar ones: these solar noises cannot offer a clean neutrino filter, but they may be soon (before lunar ones) discovered 
We do not want to hide here that the most important and (we believe at hand) tau neutrino astronomy is based on its PeV (or tens hundred PeV)  showering from mountain or valley or better via the Earth crust, in  upward air-shower toward top mountain array \cite{Fargion(2002)}.\cite{Fargion_2004} or array detector in balloons or satellites. However the ability of this modest thin and far Moon shadows to select and amplify muon (and tau) peculiar neutrino signatures and their ramified tracks it is very surprising and fashionable. In conclusion these signals might be soon or later observable by CTA (or by LHAASO)  if the astrophysical neutrino spectra mimic a  very soft (or mild soft)  spectra ruled by a very negative index -3$\div\,$-2.7; otherwise, for a hard astrophysical spectra index -2 the $\tau$ role the detection will arise at a few or tens of PeVs gamma-hadron air-showers better observable also by future GRAND \cite{2016_GRAND} array toward the very dark CR Moon sky.

\section*{Acknowledgements}
The work by MK was supported by Russian Science Foundation and fulfilled in the framework of MEPhI Academic Excellence Project (contract n. 02.a03.21.0005, 27.08.2013).


\end{document}